# Decisive role of magnetic degrees of freedom in a scenario of phase transformations in steel

I. K. Razumov[1,2*], D. V. Boukhvalov[3], M. V. Petrik[2], V. N. Urtsev[4], A. V. Shmakov[4], M. I. Katsnelson[5,6], and Yu. N. Gornostyrev[1,2]

[1]Institute of Quantum Materials Science, Ekaterinburg, 620075, Russia
[2]Institute of Metal Physics, Russian Academy of Sciences-Ural Division, Ekaterinburg, 620041, Russia.
[3]School of Computational Sciences, Korea Institute for Advanced Study (KIAS) Hoegiro 87, Dongdaemun-Gu, Seoul, 130-722, Republic of Korea
[4]Research and Technological Center Ausferr, Magnitogorsk, 455000, Russia
[5]Radboud University Nijmegen, Institute for Molecules and Materials, Heyendaalseweg 135, Nijmegen, 6525AJ, Netherlands
[6]Dept. of Theoretical Physics and Applied Mathematics, Ural Federal University, Mira str. 19, Ekaterinburg, 620002, Russia

## Abstract

Diversity of mesostructures formed in steel at cooling from high temperature austenite (γ) phase is determined by interplay of shear reconstructions of crystal lattice and diffusion of carbon. Combining first-principle calculations with large-scale phase-field simulations we demonstrate a decisive role of magnetic degrees of freedom in the formation of energy relief along the Bain path of γ-α transformation and, thus, in this interplay. We show that there is the main factor, namely, magnetic state of iron and its evolution with temperature which controls the change in character of the transformation. Based on the computational results we propose a simple model which reproduces, in a good agreement with experiment, the most important curves of the phase transformation in Fe-C, namely, the lines relevant to a start of ferrite, bainite, and martensite transformations. Phase field simulations within the model describe qualitatively typical patterns at these transformations.

PACS: 63.70.+h, 64.60.qe, 75.50.Bb, 71.15.Nc

## 1. Introduction

Despite a broad distribution of numerous new materials steel known from ancient times remains the main construction material of our civilization [1], due to high availability of its main components (Fe and C) and diversity of properties reached by a realization of various (meso)structural states [2,3]. One can control the structural state of steel due to a rich phase diagram of iron with several structural transformations at cooling from moderately high temperatures ($\delta \to \gamma \to \alpha$); the presence of carbon adds carbide phases, cementite $Fe_3C$ being the most important one. Development of the phase transformations in steel includes two main types of processes, the crystal lattice reconstruction and redistribution of carbon between the phases. Depending on the rates of these processes metallurgists separate three main types of the transformations, namely, ferrite, bainite, and martensite [2,3,4]. In practice, all transformations (except the martensitic one) involve both shear and diffusion mechanisms, their relative importance being changed with the temperature increase [4]. The difference between these types of transformations determines the diversity of properties of steel and therefore is of crucial importance for our understanding of metallurgical processes. However, there is still no commonly accepted quantitative theory which could describe the change of transformation mechanism with temperature from martensitic (lattice instability) to ferrite (nucleation and growth).

Here we demonstrate that the main factor determining scenario of the phase transformations in steel is the magnetic state of Fe and its temperature dependence. Empirically, the temperature of $\gamma$-$\alpha$ transformation in elemental Fe is close to the Curie temperature of $\alpha$-Fe; therefore the idea on the decisive role of magnetism in phase transformations for pure iron looks natural and was discussed many times; for review, see Ref. [5].

Based on state-of-the-art first-principle calculations and combining it with the phase field simulations [6] we build a consistent model which allows us to estimate (with a surprisingly high accuracy, keeping in mind its simplicity) temperature ranges corresponding to the three types of the transformations. This model includes a generalized Ginzburg-Landau functional for the Bain transformation path with ab initio parameterization and nonlinear elasticity equations for the tetragonal deformation, as well as diffusion equation for the carbon concentration. Therefore it takes into account both carbon diffusion and lattice and magnetic degrees of freedom of iron.

## 2. Methods

### 2.1. Generalized Ginzburg-Landau functional for the Bain transformation path

The minimal set of variables which is necessary to describe the $\gamma$-$\alpha$ transformation in steel includes Bain tetragonal deformation and carbon concentration. Other relevant degrees of freedom are volume per atom and magnitude of magnetic moment but we assume (following Ref. [5]) that they are fast and can therefore be taken into account just by optimization of the total energy along the Bain transformation path. The parameter of short-range magnetic order is introduced as for the case of pure iron [5].

A generalized Ginzburg-Landau functional for the total energy can be represented in the form [7]:

$$G = \int \left( g_e + \frac{k_t}{2}(\nabla e_t)^2 \right) dr, \quad (1)$$

where $g_e$ is the energy density of lattice deformations, $k_t$ is a parameter determining the width of interphase boundary [7]. We restrict ourselves by a two-dimensional model when $g_e$ can be chosen as [8,9]: $e_v = (\varepsilon_{xx} + \varepsilon_{yy})/\sqrt{2}$

$$g_e = g_t(e_t, c, T) + \frac{A_v}{2} e_v^2 + \frac{A_s}{2} e_s^2, \quad (2)$$

where $e_v = (\varepsilon_{xx} + \varepsilon_{yy})/\sqrt{2}$ is dilatation, $e_t = (\varepsilon_{xx} - \varepsilon_{yy})/\sqrt{2}$ tetragonal deformation, $e_s = \varepsilon_{xy}$ shear deformation, and $g_t(e_t, c, T)$ is the energy density depending on the tetragonal deformation parameter, local carbon concentration, and temperature. Using two-dimensional model is, of course, a simplification which does not provide the complete picture of morphology after transfromation since we have two orientation options for α-phase. Nevertheless, this model gives correctly thermodynamic condition of transformation and describes main qualitative features of microstructure formation [5,8,9]. Similar to Ref. [5] we assume that in $\gamma$-phase (initial phase for the transformation) $e_t = 0$ and in $\alpha$-phase $e_t = 1 - 1/\sqrt{2}$. The coefficients $A_v, A_s$ are expressed via elastic moduli [7], $A_v = C_{11} + C_{12}$, $A_s = 4C_{44}$. Following Ref. [5] we determine the energy density of tetragonal deformation as

$$g_t(e_t, c, T) = g_{PM}(e_t, c) - \tilde{J}(e_t, c) Q(T) \qquad (3)$$

where $\tilde{J}(e_t, c) = m^2 z_1 J_1 / \Omega = g_{PM}(e_t, c) - g_{FM}(e_t, c)$ is exchange energy, $\Omega$ is the volume per atom, $z_1$ is the nearest-neighbor number, $J_1$ is the exchange integral, $m$ is the magnetic moment, and

$$Q(T) \equiv <\mathbf{m}_0 \cdot \mathbf{m}_1> / m^2 \qquad (4)$$

is the spin correlation function dependent on temperature. We have improved our model for the temperature dependence of the nearest-spin correlator $Q(T)$ in comparison with our previous work [5]. Namely, we use Oguchi model [10] and determine $Q(T)$ as $Q(T) \sim 1/T$ for $T>T_C$; for $T<T_C$ we use the empirical formula for magnetization [11], choosing parameters in such a way that $Q(T_C) \sim 0.4$, according to Ref. [10]. Thus, at $T=T_C$ the dependence $Q(T)$ has a cusp. Curie temperature $T_C$ is related to the exchange parameter as $kT_C(e_t) = \lambda \tilde{J}(e_t) \Omega$, with the numerical factor for $\alpha$-Fe $\lambda_\alpha = 0.472$; this choice of $\lambda_\alpha$ provides an agreement of the Curie temperature with the experiment, $T_C$=1043K. The correlator for $\gamma$-Fe is chosen in a similar way, with the Curie temperature $T_C^{fcc} \approx 300$K, according to the calculations [12] for the atomic volume $\Omega \approx 12 \text{Å}^3$; $\lambda_\gamma = 0.606$ according to Ref. [13]. The temperature dependences of the correlators are shown in Figure 1.

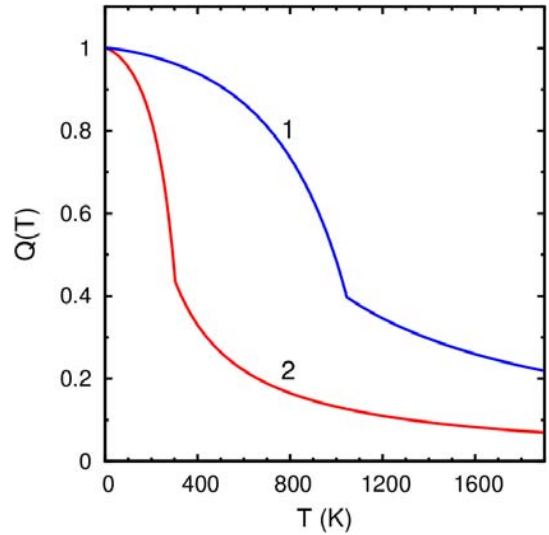

**Figure 1.** Temperature dependences of the spin correlator $Q(T)$ for $\alpha$-Fe (1) and $\gamma$-Fe (2).

To calculate free energy, we have to add entropy contributions to Eq.(1). The magnetic entropy is calculated, as in Ref. [5], from Hellmann-Feynman theorem. The configurational entropy of carbon is found from the model of ideal solutions, assuming that for $T > 300K$ carbon is equally distributed among all three interstitial sublattices in $\alpha$-Fe whereas in $\gamma$-Fe carbon atoms can occupy only quarter of the interstitial positions [14]. As a result, we obtain the following formula for the local density of free energy:

$$f(c,e_t,T) = g_{PM} - Ts_0 f_s(e_t) - \int_0^{\tilde{J}} Q(\tilde{J}',T)d\tilde{J}' + kT\left[c\ln(4c) + \left(c\ln\frac{c}{3} - c\ln(4c)\right)(1-f_s(e_t))\right] \quad (5)$$

Here $f_s(e_t)$ is a function provided a gradual switching of the entropy contribution from fcc to bcc ($f_s(e_t)=1$ in fcc and $f_s(e_t)=0$ in bcc phase); $s_0$ is the high-temperature limit of the entropy difference between the phases including phonon contribution. It is commonly accepted (see, e.g., Ref. [15]) that the value $s_0$ is almost temperature independent at $T>T_D$, where $T_D$ is Debye temperature (equal to 473K and 324 K for bcc and fcc phases, respectively). We will assume that it is a constant. The latter has been chosen such that the start of the transformation determined by the condition $\Delta f(T) = f^{fcc}(T) - f^{bcc}(T) \equiv 0$ agrees with the experimental value for elemental Fe, $T_0 = 1184K$. This gives us the value $s_0 = -0.19k$, quite close to the experimental estimate [16].

The resulting Ginzburg-Landau functional for the free energy reads:

$$F = \int\left(f(c,e_t,T) + \frac{A_v}{2}e_v^2 + \frac{A_s}{2}e_s^2 + \frac{k_t}{2}(\nabla e_t)^2\right)dr \quad (6)$$

The quantities $g_{PM}(e_t,c), g_{FM}(e_t,c)$ are found from the energy curves along the Bain path for para- and ferromagnetic states, respectively. Carbon shifts the thermodynamic potentials of γ and α phases of Fe in accordance with its solution enthalpy. As was shown in Ref. [17], carbon turns out to effect dramatically on magnetic state of γ-Fe; it can create a locally ferromagnetically polarized region with tetragonal distortions. Thus, thermodynamics of γ-Fe-C system, in particular, solution enthalpy of carbon, should be strongly dependent on local magnetic order. Here we include the dependence of the energies of γ and α phases on carbon concentration into the model based on first-principles electronic structure calculations of the solution enthalpy.

## 2.2. First-principle calculations

The calculations of energetics of Fe-C system were performed by density functional theory in the pseudopotential code SIESTA [18], similar to our previous work [17]. All calculations were carried out using the generalized gradient approximation (GGA-PBE) with spin-polarization [19]. Full optimization of the atomic positions was performed. During the optimization, the ion cores were described by norm-conserving pseudo-potentials [20] and the wave functions were expanded with a double-ζ plus polarization basis of localized orbitals for iron and carbon. Optimization of the forces and total energy was performed with an accuracy of 0.04 eV/Å and 1 meV, respectively. All calculations were carried out with an energy mesh cut-off of 300 Ry and a k-point mesh of 4×4×4 in the Mokhorst-Park scheme [21]. For the modeling of all configurations 3×3×3 supercell of 108 iron atoms in fcc configuration was used. Varying of the concentration of carbon was realized by the change of the number of interstitial carbon atoms in the voids from one (~1 at%) to three (~3 at%). For the modeling of paramagnetic configuration five possible special quasi-random structures (SQS [22]) of magnetic moments were generated by reinitialize each time the pseudo-random number generator. The structure with the lowest total energy have been defined as a ground state and energy difference per iron atoms have been used to estimate the error of the modeling of paramagnetic iron. The modeling of the Bain pathways was performed by the method previously employed for the pure iron [12]. In contrast to Ref. [12], to take into account thermal expansion effects the elementary cell volume was chosen close to experimental values for $\gamma$- and $\alpha$-Fe at the temperature of γ-α transition and linearly interpolated for $1/\sqrt{2} < c/a < 1$ (actually, the change of the lattice constant along the path is within 1%). The difference of the energies between ferromagnetic and paramagnetic state agrees well with the "exchange energy" calculated in Ref. [12], thus, the different choice of the lattice constant is not essential.

The energies found from the first-principle calculations for pure iron were approximated by the following polynoms:

$$\widetilde{g}_{PM(FM)}(\phi) = g_{PM(FM)}^{fcc} + 2\left(g_{PM(FM)}^{bcc} - g_{PM(FM)}^{fcc} + \frac{c_{PM(FM)}}{6}\right)\left(\phi^2 - \frac{\phi^4}{2}\right) + c_{PM(FM)}\left(\frac{\phi^6}{3} - \frac{\phi^4}{2}\right) \quad (7)$$

Here the order parameter $-1 < \phi < 1$ related to the Bain tetragonal deformation as $\phi = \sqrt{2}/(\sqrt{2}-1)e_t$. Positive and negative values of $\phi$ correspond to two possible (mutually orthogonal) directions of the Bain deformation in two-dimensional case. Its form guarantees extrema at the points $\phi = 0$ or $\phi = \pm 1$, parameters $g_{PM}^{fcc}$, $g_{FM}^{fcc}$, $g_{PM}^{bcc}$, $g_{FM}^{bcc}$, $c_{PM}$, $c_{FM}$ were found by fitting to the ab initio computational results.

We do not take into account carbon-carbon interactions, due to a smallness of carbon concentration. Thus, its contribution was taken as linear:

$$g_{PM(FM)}(\phi,c) = \widetilde{g}_{PM(FM)}(\phi) + c\varepsilon_{PM(FM)}^{fcc} + c\left(\varepsilon_{PM(FM)}^{bcc} - \varepsilon_{PM(FM)}^{fcc}\right)(1 - f_s(\phi)) \quad (8)$$

Function $f_s(\phi)$ have been chosen in form $f_s(\phi) = (1 - \phi^2)^2$. Within the approximation (8) the effect of carbon on Bain-path energetics is determined only by carbon solution energies in $\gamma$- and $\alpha$- phases. We deal with the temperatures $T > 400K$ where carbon fills equally all three sublattices of octahedral interstitials and therefore we do not takes into account tetragonality of martensite which arises at $T \approx 300K$ [14].

Parameterization of these formulas from ab initio calculations leads to the following values: $g_{PM}^{bcc} = 0.19$, $g_{PM}^{fcc} = 0.14$, $g_{FM}^{fcc} = 0.095$, $g_{FM}^{bcc} = 0$ (in eV/at) and $c_{PM} = 0.05$, $c_{FM} = -0.08$ (all in eV/at). These data were slightly different from those calculated by us earlier [12] by VASP (the energy $g_{PM}^{bcc}$ coincides with Ref. [12], the energy $g_{PM}^{fcc}$ differs by -0.02eV/at). The solution energies of carbon in different phases, $\varepsilon_{FM}^{bcc} = 0.8$, $\varepsilon_{PM}^{bcc} = 0.7$, $\varepsilon_{FM}^{fcc} = -0.2$, $\varepsilon_{PM}^{fcc} = 0.22$ (in eV/at) were chosen on the base of similar calculations for iron with carbon concentration ~1% at. The value $\varepsilon_{FM}^{bcc}$ agrees with the result of the previous first-principle calculations [23]; $\varepsilon_{PM}^{fcc}$ agrees with the result [17], but lower than the experimental value 0.4eV/at [24].

### 2.3. Kinetic equations

It was shown in Refs. [7,25] that at the description of atomic displacements in solids one cannot take into account only the order-parameter (in our case, tetragonal deformation) since other components of the deformation tensor are coupled to the order parameter by Saint Venant compatibility equations. The latter result in effective long-range interactions which are crucial for the pattern formation at the transition [6,8,25]. Therefore, following Ref. [6] we write the dynamical equations for atomic displacements in a form similar to Newton equations rather than Allen-Cahn time dependent Ginzburg-Landau relaxation equation for the order parameter [26,27]. It allows taking into account automatically the Saint Venant compatibility equations.

We exploit the equations of motion used by us earlier for elemental iron [5] plus the equation of carbon diffusion:

$$\rho \frac{\partial^2 u_i(\mathbf{r},t)}{\partial t^2} = \sum_j \frac{\partial \sigma_{ij}(\mathbf{r},t)}{\partial r_j} \qquad \frac{\partial c}{\partial t} = -\nabla \mathbf{I} \qquad (9)$$

Here $\rho$ is the mass density of iron; elastic stresses $\sigma_{ij}$ and a flow of carbon atoms $I$ are calculated via variational derivatives of the Ginzburg-Landau functional:

$$\sigma_{ij}(\mathbf{r},t) = \frac{\delta F}{\delta \varepsilon_{ij}(\mathbf{r},t)} \; , \qquad \mathbf{I} = -\frac{D}{kT} c(1-c) \nabla \left( \frac{\delta F}{\delta c} \right) \qquad (10)$$

where $D$ is carbon diffusion coefficient (see Appendix); the deformations $\varepsilon_{ij}$ introduced above are connected with the variable of our model as

$$\phi = \sqrt{2}/(\sqrt{2}-1) e_t, \qquad e_v = (\varepsilon_{xx} + \varepsilon_{yy})/\sqrt{2}, \qquad e_t = (\varepsilon_{xx} - \varepsilon_{yy})/\sqrt{2} \qquad (11)$$

$$\varepsilon_{xx} = \frac{\partial u_x}{\partial x}, \quad \varepsilon_{yy} = \frac{\partial u_y}{\partial y}, \quad \varepsilon_{xy} = 0.5 \left( \frac{\partial u_x}{\partial y} + \frac{\partial u_y}{\partial x} \right) \qquad (12)$$

As a result,

$$\sigma_{xx} = \frac{1}{(\sqrt{2}-1)} \frac{df(c,\phi,T)}{d\phi} + \tilde{A}_v e_v - \tilde{k}_t \nabla^2 \phi, \qquad (13)$$

$$\sigma_{yy} = -\frac{1}{(\sqrt{2}-1)} \frac{df(c,\phi,T)}{d\phi} + \tilde{A}_v e_v + \tilde{k}_t \nabla^2 \phi, \qquad (14)$$

$$\sigma_{xy} = A_s e_s \qquad (15)$$

where $\tilde{A}_v = A_v/\sqrt{2}$, $\tilde{k}_t = k_t(\sqrt{2}-1)/2$, $k_t = 10^{-3}$ (in the units of $L^2 \Omega \tilde{J}^\alpha$ where $L$ is the sample size, $\Omega \tilde{J}^\alpha = 0.19$ eV/at). We pass further to dimensionless units $r_i \to r_i/L$, $u_i \to u_i/L$, $t \to t \sqrt{\frac{\tilde{J}^\alpha}{L^2 \rho}}$, $\rho \to 1$, $D \to D \sqrt{\frac{\rho}{L^2 \tilde{J}^\alpha}}$, $\sigma_{ij} \to \sigma_{ij}/\tilde{J}^\alpha$.

To maximize the size of the system under simulation for given computer resources we restrict ourselves to the two-dimensional case. It is enough to distinguish clearly patterns typical for different transformations in Fe-C. The details of the simulations are presented in the Appendix.

### 3. Results and discussion

#### 3.1. Bain path and free energy in Fe-C

The Bain path is the tetragonal deformation accomplished $\gamma$-$\alpha$ lattice reconstruction, which change from $c/a = 1$ for fcc ($\gamma$) to $c/a = 1/\sqrt{2}$ for bcc ($\alpha$) structures. Dependences of the energy (*g*) and free energy (*f*) per Fe atom on tetragonal distortion calculated according to the formulas (3–5), (7–8) are shown in Figure 2. One can see that the ratio of the energies for α and γ phases changes strongly with the temperature decrease and *α*-phase becomes preferable at $T < T_C$. Figure 3 displays the temperature dependence of the energy $\Delta g(T) = g^{fcc}(T) - g^{bcc}(T)$ and free energy difference $\Delta f(T) = f^{fcc}(T) - f^{bcc}(T)$ in comparison with the data [16] for elemental iron. One can see on this figure that the model constructed with correlator $Q(T)$ (see Figure 1) describes correctly thermodynamics of both phases of pure Fe within the temperature range 600÷1200K and agrees well with the results of CALPHAD [16]. It turns out that the magnetic contribution dominates at $T \leq T_C$ and is compensated essentially by the phonon contribution at $T > T_C$.

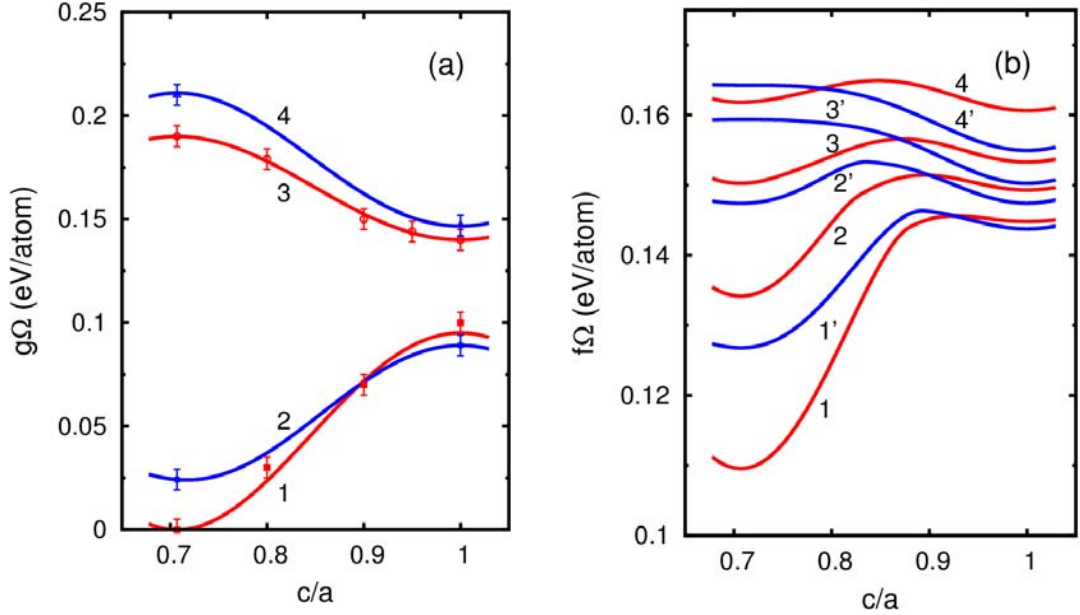

**Figure 2.** (Color online) Energy (a) resulting from the first-principle calculation for the Bain path in ferro- (curves 1,2) and paramagnetic (3,4) states for carbon concentration *C* = 0 (1,3) and *C* = 3 at% (2,4). Free energy (b) as functions of tetragonal deformation for temperatures *T*=600K (curves 1,1'), 800K (2,2'), 1000K (3,3'), 1400K (4,4') found from Eqs. (5) and the first-principle computational results for carbon concentration *C* = 0 and *C* = 3 at%, respectively. Symbols correspond to the computational results, solid lines are approximations used in the model.

In elemental Fe, for ferromagnetic state γ-phase corresponds to the maximum of the total energy, instead of local minimum and therefore the transition to α-phase happens without barrier [28]. It turns out that doping by carbon does not change this important peculiarity. Moreover, carbon decreases the energy of ferromagnetic γ-Fe, with the enthalpy solution of the order of -0.2 eV per carbon atom (Figure 2). It is not surprising since carbon creates a strong local ferromagnetic order in paramagnetic or antiferromagnetic γ-Fe [17]. For the other cases (*α*-phase and paramagnetic γ-Fe) the solution enthalpy of carbon is positive. It is a common wisdom that interstitial impurities (including carbon) always prefer fcc surrounding compared to bcc, just for geometric reasons (the voids are larger in fcc lattice than in bcc with the same density) [29]. This is for sure correct, also for carbon in iron and results in a more pronounced effect of carbon addition on energy bcc-Fe. What is much less trivial is that carbon solubility in fcc iron is very sensitive to the magnetic state being maximal in ferromagnetic surrounding.

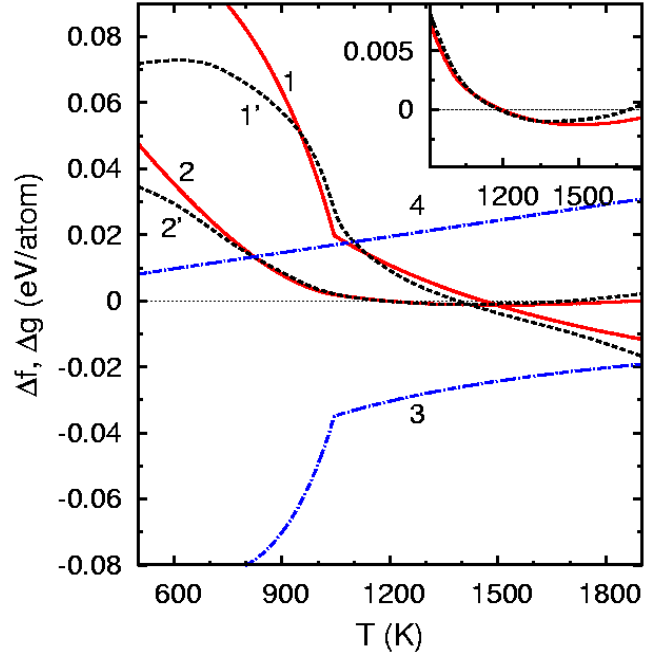

**Figure 3.** The energy difference $\Delta g(T) = g^{fcc}(T) - g^{bcc}(T)$ (curve 1) and free energy difference $\Delta f(T) = f^{fcc}(T) - f^{bcc}(T)$ (curve 2) at $\gamma \rightarrow \alpha$ transition in elemental iron in comparison with known data (dotted lines 1',2') [16]; contribution of magnetic entropy to the free energy (curve 3) and the contribution from phonon entropy (curve 4).

### 3.2. Construction of the phase diagram and scenarios of transformations in steel

Now we are ready to discuss the difference between scenarios of phase transformations in our model. This difference originates from the strong temperature dependence of driving force for the transformation, the rate of carbon diffusion and plastic relaxation of transformation strain. As discussed above, the strong temperature dependence of the former (followed from the strong temperature dependence of the potential transformation relief) is magnetic in origin: the temperature enters our model mainly via the parameter of a short-range magnetic order.

Ferrite transformation kinetics is controlled by carbon diffusion. Without the redistribution of carbon, α-phase is not thermodynamically favorable and therefore ferrite formed by the mechanism of heterogeneous nucleation, usually at grain boundaries. At a low enough overcooling below the temperature $A_3$ [2,3], determined by the condition of equality of chemical potentials for $\alpha$-phase depleted by carbon and $\gamma$-phase enriched by carbon, and restricting the two-phase $\gamma+\alpha$ region, the ferrite transformation proceeds slowly since its driving force is small and a realization of the transformation requires a redistribution of carbon at large distances. Thermodynamic potentials of $\alpha$-phase without carbon and $\gamma$-phase with nominal carbon concentration are equalized at a temperature $T_F < A_3$, when $f(e_t^\gamma, c_0, T) = f(e_t^\alpha, c = 0, T)$, $c_0$ is initial (average over the sample) carbon concentration. One can expect that at $T \leq T_F$ the $\gamma$-$\alpha$ transformation accelerates essentially since in this case the short-range carbon diffusion is sufficient. Therefore we identify the temperature $T_F$ with the start of rapid ferrite transformation. It should be noted that $T_F$ appears to be close to Curie temperature $T_C$ in a broad range of carbon concentration.

Further decrease of temperature results in a slowdown of carbon diffusion and enhancement of the transformation driving force. At intermediate temperatures, a crucial role in determining of the temperature of start of transformation [4] is played by a temperature of paraequilibrium $T_0$ where the free energies of $\alpha$- and $\gamma$-phases with the same carbon concentration become equal, $f(e_t^\gamma, c_0, T) = f(e_t^\alpha, c_0, T)$. Temperature $T_0$ was introduced in Ref. [30] as a pre-condition for the start of bainite transformation. In this case, as it assumed in [4,30], the diffusion is slower than the shear transformation and therefore there is no redistribution of carbon between $\alpha$- and $\gamma$-phases during the growth of $\alpha$-phase plates. The value of $A_3$ and $T_0$ calculated by us agrees well with the experimental quantity $A_3^{\exp}$ and $T_{0Z}$ (Figure 4).

At last, the martensite transformation is characterized by mechanical instability of $\gamma$-phase with carbon, that is, the free energy as a function of tetragonal deformation should have a maximum instead of minimum at the fcc point, $\partial^2 f(e_t, c, T)/\partial e_t^2 = 0$. This condition is attained by quenching of $\gamma$-phase to the temperature $M_S$ where ferromagnetic short range order in $\gamma$-phase becomes important. One can see that the temperature $M_S$ found in this way is actually lower than the experimental value (see Figure 4). One has to keep in mind, however, that the martensitic transformation observed in steel do not follow the scenario of lattice instability and is developed, rather, by heterogeneous nucleation and "replication" mechanism discussed previously [5]. Indeed, it was shown in Ref. [5] that above $M_S$ a broad temperature range exists where the transformation is martensite-like but includes nucleation and growth processes. We follow the concept of isothermal martensitic transformation [31–34] and accept the condition of martensite start as $f_{barrier}^{\gamma \to \alpha} = C_0 kT$ where parameter $C_0$=0.04 is chosen by fitting to the experiment for pure Fe [35]. The temperature $M_{S'}$ determined in this way agrees well with the experiment in a broad interval of carbon concentration.

With these definitions, the curves $A_3, T_0, T_F$ do not depend on the energy relief along the Bain path and are determined only by terminal values $g_{PM}^{fcc}$, $g_{FM}^{fcc}$, $g_{PM}^{bcc}$, $g_{FM}^{bcc}$. Contrary, the martensitic curves $M_{S'}, M_S$ depend on the energetics at intermediate $e_t$. For the concentration range under consideration the magnetic order effects in $\gamma$-Fe are negligible, for the temperatures above $T$~ 400K. Therefore the general shape of the phase diagram (the lines $A_3$, $T_f$, $T_0$, $M_{S'}$) are determined, first of all, by the evolution of magnetic state in $\alpha$-Fe. In particular, the $\gamma \to \alpha$ transition turns out to be possible above Curie temperature ($T_C^{bcc} \approx$1043K) due to the short-range ferromagnetic order in $\alpha$-Fe (see also Ref. [5]). The short range magnetic order in $\gamma$-Fe becomes important at $T \approx$ 400K, which determines the temperature of start of the martensitic transformation $M_S$, developing via the lattice instability.

The results presented in Figure 4 are purely thermodynamic for the lines $A_3, T_0, T_F$ and do not take into account the internal strain produced by transformation which plays a crucial role in phase morphology and transformation kinetics. Due to requirements following from Saint Venant compatibility equations, the resulting Ginzburg-Landau functional for the free energy should include different components of the deformation tensor as well as their gradients (6). Besides, the plastic relaxation of elastic stresses accompanying the formation of the new phase is another important factor which was taken into account in a model way (see Appendix). We use the phase-field model formulated earlier for the elemental iron [5], generalizing it with taking into account diffusive redistribution of carbon. Therefore, we describe the transformation kinetics by equations for atomic displacements, plus diffusion equation for carbon, using the Ginzburg-Landau functional (6), see Methods section for details.

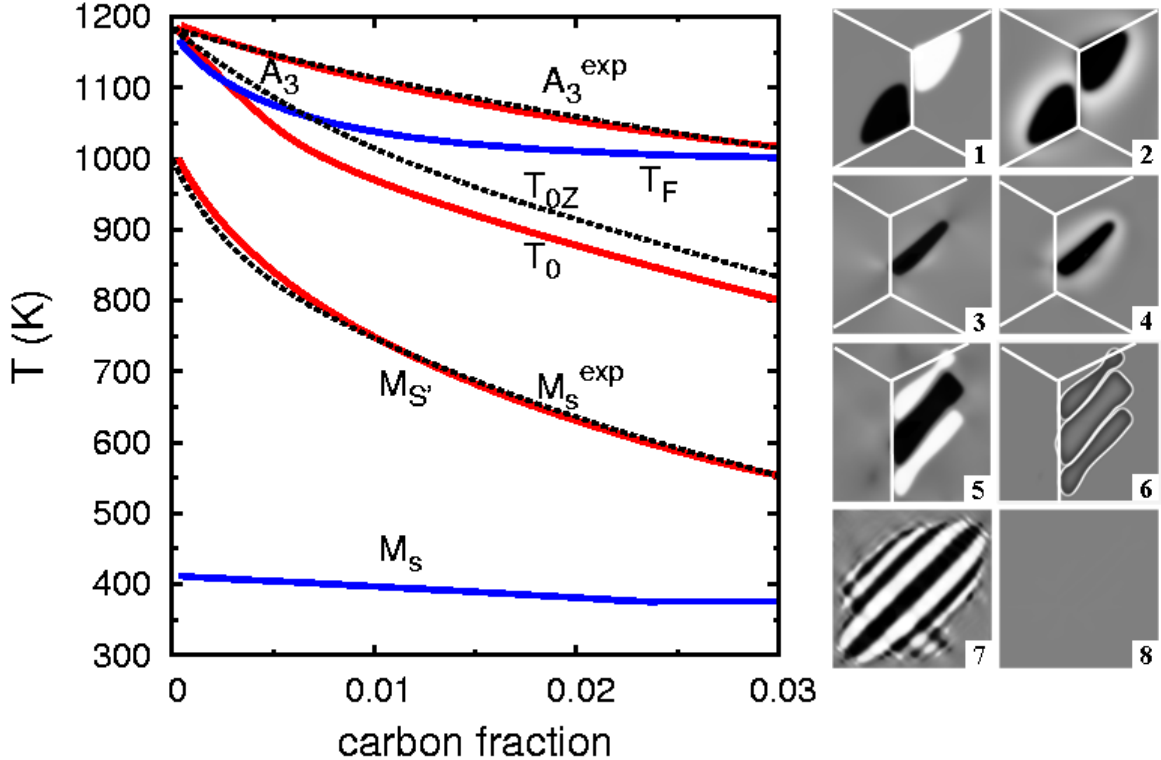

**Figure 4.** The left panel shows calculated lines (solid) corresponding to the start of ferrite transformation, paraequilibrium, and the start of martensitic transformation. $M_s$ and $M_{s'}$ are temperatures start of lattice instability and martensitic-like transformation. Dashed lines show experimental boundary of two-phase region ($A_3$) [36], experimental paraequilibrium temperature ($T_{0Z}$) [37], and experimental temperature of start of martensitic transformation ($M_s^{\exp}$) [35]. The right panel shows microstructures forming as a result of transformation at various temperatures: $T_0<T<A_3$ (1,2), $M_{s'}<T<T_0$ (3,4; 5,6), $T<M_{s'}$ (7,8). The left and right columns at this panel correspond to tetragonal strain (black and white are two orthogonal directions of tetragonal deformation in bcc phase, grey shows fcc regions) and carbon distribution (the darker the smaller), respectively.

In the right side of Figure 4 we show typical patterns of tetragonal deformation (first column) and carbon distribution (second column) obtained in our phase field simulations for different temperatures. The ferrite transformation starts as a heterogeneous nucleation of $\alpha$-phase. One can see, indeed, that at the temperature $T<A_3$ carbon leaves $\alpha$-phase and this process controls a formation of the new phase. For this situation we see arising polygonal particles of carbon-free $\alpha$-phase surrounded by carbon-reach shell that are really typical for ferrite transformation [3].

For the temperature range $M_{S'}<T<T_0$ the model demonstrates several possible scenarios. At small cooling below $T_0$ the transformation develops by the diffusion mechanism. In this case, due to incomplete relaxation of the internal stresses $\alpha$-phase has a shape of plate. For deeper cooling a fast growth of the plate via the shift mechanism is possible and redistribution of carbon between $\alpha$- and $\gamma$-phases happens only after the plates are formed. Most of carbon atoms sit at the interphases of plates with different orientations. All these features are indeed characteristic of early stage of bainite transformation [2,4]. For a longer exposure time the formation of the cementite particles takes place that is beyond a scope of our consideration. Finally, at $T<M_{S'}$ local fluctuations initiate martensite transformation which results in a formation of lenticular colony of tweens with carbon homogeneously distributed over the sample.

## 4. Discussion

To conclude, we propose a microscopic model describing, in agreement with experiment, the curves at Fe-C phase diagram separating regions of ferrite, paraequilibrium (bainite), and martensite transformations. We were able not only describe these phenomena but, to some extent, understand them separating the main factor, namely, the temperature dependence of magnetic short-range order. The curves of start of ferrite, paraequilibrium (bainite), and martensite transformations are shown in Figure 4, together with known experimental data. This is the main result of our work. Keeping in mind that our model is ab-initio based (does not contain fitting parameters except the threshold value of energy barrier for martensitic transformation) one can consider the agreement as amazingly good. One should stress that this agreement is reached for the model where main temperature dependence enters via the degree of short-range magnetic order. Thus, the closeness of the Curie temperature in bcc iron to the temperature of structural transformation is not accidental but is related with the essence of phase transformations both in elemental iron [5] and steel.

## Appendix. Simulation of transformation kinetics

Relaxation processes of elastic fields play an essential role in transformation kinetics and morphology of the new phase. The main channel of such a relaxation is a plastic deformation arising when the stresses exceed the yield stress. A consequent description of the plastic deformation requires an essential complication of the model, by adding parameters describing the plastic deformation to the corresponding dynamic equations [38,39]. Instead, we take into account the plastic deformation in a phenomenological way. Since the contribution of the elastic stresses to the Ginzburg-Landau functional is determined by the coefficients $A_v, A_s$, we replace the real values of these parameters by some effective, temperature dependent values. The scheme proposed provides the stress relaxation assuming that the relaxation processes are faster than typical times of development of the transformation and that the lattice remains coherent during the whole process. We assume that for ferritic temperatures ($T>T_0$) where the transformation velocity is limited by the carbon diffusion the stresses have enough of time to relax completely, choosing therefore $A_v^{eff} = A_s^{eff} \approx 0$. Contrary, the martensitic transformation occurs with the velocities comparable with the speed of sound and therefore for $T < M_S$ there is no relaxation within the relevant time interval, therefore $A_v^{eff} = A_v$, $A_s^{eff} = A_s$. The values of the parameters $A_v$, $A_s$ where chosen as in Ref. [5]. For the temperature range $M_S < T < T_0$ intermediate values of the coefficients $A_v^{eff}$, $A_s^{eff}$ were used (see Table 1, lines I and II). Parameters $A_v^{eff}$, $A_s^{eff}$ for martensite (Figure 4, fragments numbered 7 and 8) were chosen in such a way that the average elastic energy over the sample was equal to the experimental value of the stored energy in martensite, 0.007eV/at [40]. For the other temperature ranges these parameters were chosen according to the experimentally known values of the stored energies for Widmanstaetten ferrite, bainite and martensite [41].

The diffusion coefficient of carbon $D$ is different in $\alpha$- and $\gamma$-phases and temperature dependent. We use a simple expression $D = D_\gamma + (D_\alpha - D_\gamma)\phi^2(2-\phi^2)$, where $D_\alpha$, $D_\gamma$ are handbook data [42], for which we use approximations (m²/s): $D_\gamma = 4.5 \cdot 10^{-5} \exp(-18530/T)$, $\lg D_\alpha = -4.9 - 0.52X + 1.61 \cdot 10^{-3} X^2$, $X = 10^4/T$. In particular, at $T$=1000K the ratio $D_\alpha/D_\gamma \approx 300$, that is, at the precipitation of $\alpha$-phase carbon is expelled into the boundary layer but only weakly diffuse into the bulk of $\gamma$-phase.

We do not take into account temperature-induced lattice fluctuations. The latter are mostly important for homogeneous nucleation whereas we deal with inhomogeneous nucleation at grain boundaries. Indeed, it is known experimentally that ferrite nucleates preferably at grain boundaries and their triple joints. To describe this process we consider a region with two triple joints of grains and introduce an additional contribution to the free energy near the grain boundary,

$$\Delta f_{GB}(x) = \Delta f_{GB}^0 \phi^2(2-\phi^2)P(x), \qquad P(x) = \frac{4\sqrt[4]{3}}{3}\frac{\lambda x}{1+(\lambda x)^4} \qquad (16)$$

where $x$ is the distance from the grain boundary (in dimensionless units as described above) in the direction perpendicular to the boundary, $\Delta f_{GB}^0$ is the maximal amplitude of the perturbation, $\lambda$ is the parameter characterizing the width of the grain boundary. This means that a near-boundary region is favorable for the transformation but its penetration through the boundary is suppressed by the change of crystal lattice orientation. Apart from this, we use the local perturbation initiating the start of the transformation as $\Delta f_{loc}(r) = \Delta f_{loc}^0 \phi/(1+(\lambda r)^6)$, where $r$ is the distance from the center of perturbation region.

The phase field simulations show that the ferrite transformation observed in the temperature range $T_0 < T < A_3$ is controlled by the diffusion of carbon and requires an essential stress relaxation; for homogeneous distribution of carbon and without stress relaxation the ferrite embryos has no thermodynamic motivation to grow. In this case, we restrict ourselves by the consideration of diffusive kinetics only and calculate the distribution of deformations from quasistationary equations $\sum_j \frac{\partial \sigma_{ij}(\mathbf{r},t)}{\partial r_j} = 0$ for a given (time-dependent) carbon distribution (Figure 4, fragments numbered 1 and 2). Since $D_\alpha \gg D_\gamma$, a carbon shell is formed around precipitates of $\alpha$-phase during the transformation.

For the temperatures $T<T_0$ the transformation can proceed even for homogeneous distribution of carbon. However, to find the temperature $T_0$ from thermodynamic condition $f(c,\phi=0,T_0) = f(c,\phi=1,T_0)$ is not enough since this condition does not take into account the contribution of elastic stresses to the free energy. Our simulations show that the stresses shift the start of the shear transformation towards lower temperatures $T<T_0$. The transformation scenario is dependent on the degree of overcooling. For higher temperature when the relaxation is strong enough (the case I in Table 1) the transformation is developed similar to ferrite one; it is controlled by redistribution of carbon but α-phase has a shape of a plate similar to Widmanstaetten ferrite [41,43] (Figure 4, fragments numbered 3 and 4). For lower temperatures (weaker stress relaxation, case II in Table 1), γ-α transformation starts as a shear one and is developed with a formation of one or several twinned plates depend on magnitude $\Delta f_{loc}^0$. By analogy with bainite transformation, one should expect that the plate stops its growth after reaching a critical size due to the loss of coherence at $\gamma$-$\alpha$ interface after a plastic deformation; further evolution is determined by diffusion of carbon, up to formation of a new plate. In this case, we perform simulations in two stages. At the first stage ($t<5$, dimensionless time was determined in section 2.3) we solve the full set of equations with real parameters $D_\alpha, D_\gamma$; only weak redistribution of carbon takes place at this stage. At the second stage ($t>5$) the distribution of deformations is frozen and only diffusive part of the dynamical problem is considered. At this stage, carbon moves from the bulk of $\alpha$-plates to the host of $\gamma$-phase. (Figure 4, fragments numbered 5 and 6).

In reality, in steel within the temperature range $M_{S'}<T<T_0$, apart from Widmanstaetten ferrite, the bainite are observed, with coexistence of shear transformation and carbon diffusion, as well as a formation of cementite [41]. To simulate the growth of bainite colony one needs to include cementite in the model and to consider in a more consistent way plastic relaxation. This issue is therefore beyond a scope of our consideration. Nevertheless, our model is applicable at the stage of nucleation and predicts two possible scenarios of the transformation. Depending on temperature, it can follow either shear or diffusive mechanisms.

|  | $T$, K | $c$ | $L$, nm | $\Delta f_{GB}^0$, eV/at | $\Delta f_{loc}^0$, eV/at | $\lambda$ | $\dfrac{A_v^{eff}}{A_v}$ | $\dfrac{A_s^{eff}}{A_s}$ | $D_\alpha$, m$^2$/s | $D_\gamma$, m$^2$/s |
|---|---|---|---|---|---|---|---|---|---|---|
| $T_0<T<A_3$ | 1000 |  |  |  |  |  | 0 | 0 | 1.2E-10 | 4.0E-13 |
| $M_{S'}<T<T_0$ (I) | 850 |  |  |  |  |  | 0.005 | 0.005 | 1.6E-11 | 1.5E-14 |
| $M_{S'}<T<T_0$ (II) | 800 | 0.01 | 500 | 0.01 | 0.03 | 50 | 0.015 | 0.015 | 7.1E-12 | 3.9E-15 |
| $M_S<T<M_{S'}$ | 700 |  |  |  |  |  | 0.050 | 0.050 | 1.1E-12 | 1.4E-16 |
| $T<M_S$ | 350 |  |  |  |  |  | 1 | 1 | 3.5E-19 | 4.6E-28 |

**Table 1.** The parameters used in the simulations.